\documentclass[journal]{IEEEtran}
\usepackage{cite}

\ifCLASSINFOpdf
\else
\fi

\usepackage{graphicx}
\usepackage{dcolumn}
\usepackage{epstopdf}
\usepackage{amsmath}
\usepackage{subfigure}
\usepackage{subfloat}
\makeatletter

\newcommand{\Rmnum}[1]{\expandafter\@slowromancap\romannumeral #1@}
\makeatother
\usepackage{color}
\usepackage{bm}
\usepackage[autostyle]{csquotes}

\begin{document}
%
\title{Periodic metallic stepped-slits for entire transmission of optical wave and efficient transmission of terahertz wave}
%
%
%

\author{Mohammadreza~Khorshidi,
        Gholamreza~Dadashzadeh
\thanks{M.~Khorshidi and G.~Dadashzadeh were with the Department
of Electrical and Electronic Engineering, Shahed University, Persian
Gulf Highway,
Tehran 3319118651, Iran.}
}

%
%

\markboth{}%
{Shell \MakeLowercase{\textit{et al.}}: Bare Demo of IEEEtran.cls for IEEE Journals}
%



\maketitle

\begin{abstract}
Metallic structures with periodic array of slits are well-known to
lead to extraordinary power transmission, when slits have dimensions
much less than the wavelength of incident optical wave. Excellent
power transmission originates from surface waves excited by incident
transverse magnetic wave. Here we show that metallic structure with
array of stepped-slits can transmit power of incident wave into the
substrate at desired optical frequency entirely and simultaneously
in terahertz frequency band by 70\%, for
$\text{In}_{0.53}\text{Ga}_{0.47}\text{As}$ substrate. Transmitted
power of the proposed structure is studied both in a closed-form by
an analytical model as well as numerically by finite element method.
It is found that with the increase of field intensity in the
substrate, structure with array of stepped-slits (as opposed to
uniform-slits) favorably has no reduction of frequency range at
maximum transmitted power.
\end{abstract}


%
\IEEEpeerreviewmaketitle

\section{Introduction}
%
%
%
%
\IEEEPARstart{S}{ince} the pioneering work of Ebbesen et
al.\cite{Ebbesen} on holes, extraordinary power transmission through
periodic metallic structures with dimensions much less than the
wavelength of incident wave, has attracted growing interests among
researchers \cite{Miyamaru,Khavasi}. This unusual at the same time
interesting phenomena, happens due to excited surface waves in
subwavelength structure which mitigate its power reflection.
Extraordinary power transmission has brought subwavelength
structures into practical applications also, in terahertz sources
and detectors \cite{Wang}, chemical sensing \cite{Yang},
spectroscopy and imaging techniques \cite{Hartschuh,Frey}, ultrafast
photodetectors \cite{Chang}, and high efficiency solar cells
\cite{Shang}, to name a few. To be more specific on the subject,
photoconductive antennas, a class of terahertz sources and
detectors, are well-known to suffer from low radiated power
\cite{Lee,Kim,Chattopadhyay,Khiabani}. Their performance can be
improved by the use of subwavelength structures which was proposed
for the first time by Jarrahi and co-workers
\cite{Berry12,Berry13,Yang14}, as a result of two facts. First, the
time required for photocarriers (electrons and holes) generated in
photoconductive area, to arrive at radiating electrodes decreases.
Second, power reflection at optical and terahertz frequencies can be
reduced, as will be shown here \cite{Khorshidi}.

In this work we propose and study the use of stepped slits for
reduction of power reflection at both optical frequency as well as
terahertz frequency band. The contents of this paper are as follows:
Section \Rmnum{2} develops a theoretical model and subsequently
obtains a closed-form expression for the power reflected from the
structure. Section \Rmnum{3} then discusses how to design a
structure with maximum power transmission at desired frequencies.
Finally, results from theoretical model are compared with finite
element method calculations for specific values of dimensions and
frequencies.
\section{Theory}
\label{sec:1} The X-Y cross view of the proposed periodic metallic
structure is shown in Fig.~\ref{fig:fig1}; all dimensions stay the
same through the Z-direction. The structure consists of two
nano-scale layered metallic rods with similar height $\xi=w_y/2$ and
different widths. The widths of the upper and bottom rods are,
respectively, $w_x$ and $w'_x$. These rods are repeated along the
X-axis with periodicity $d_x$. For the sake of generality of the
theoretical model proposed in this work for different wavelengths,
metallic rods are considered as perfect conductors with infinite
conductivity. The X-Y cross section can be seen as of consisting of
four different regions, shown by 1 through 4 in Fig.~\ref{fig:fig1}.
Regions 1 and 4 are infinite uniform mediums, and regions 2 and 3
are periodic regions. Electromagnetic wave incident to the periodic
structure can be considered as either transverse electric (TE) with
electric field or transverse magnetic (TM) with magnetic field in
the Z-direction. In this paper, the TM-polarized incident wave which
excites surface wave and subsequently contributes in extraordinary
power transmission, is considered for analysis. The theoretical
model is built up by expressing electromagnetic fields in both
uniform regions 1 and 4 as well as parallel plate waveguide (PPWG)
regions 2 (top PPWG) and 3 (bottom PPWG), as Fourier series
expansions. Scattering matrices are considered for different
interfaces existing in the structure; $\bm{S}^{(1-2)}$ for the
interface between air medium 1 and the top PPWG, $\bm{S}^{(2-3)}$
for the interface between the top and the bottom PPWGs, and
$\bm{S}^{(3-4)}$ for the interface between the bottom PPWG and the
dielectric substrate 4. In order to analyze the performance of the
proposed structure on impinging TM-polarized wave, scattering
matrices $\bm{S}^{(i-j)}$, (\textit{i}, \textit{j})=(1, 2), (2, 3),
and (3, 4), are first calculated and then combined based on cascade
networks rule as in the followings.
\subsection{Calculation of $\bm{S}^{(1-2)}$ Matrix}
\label{sec:2} To obtain the $\bm{S}^{(1-2)}$ matrix, magnetic fields
inside regions 1 and 2, respectively, are expressed as
Eqs.~(\ref{eq:one}) and (\ref{eq:two}) below (\cite{Hwang}):
\begin{figure}
\includegraphics[scale=.7]{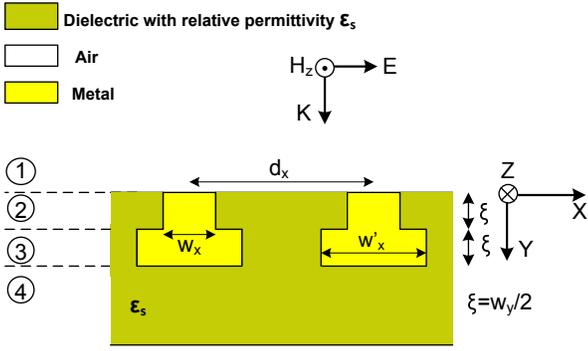}
\caption{\label{fig:fig1} X-Y cross view of the proposed
stepped-slits periodic metallic structure.}
\end{figure}
\begin{figure}
\includegraphics[scale=.4]{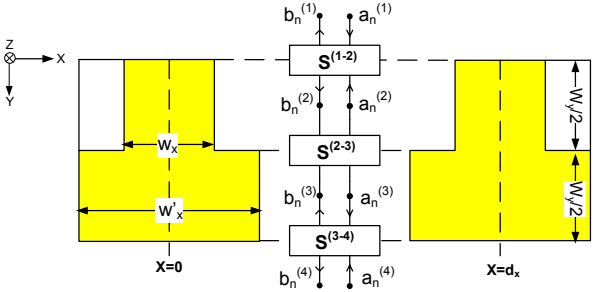}
\caption{\label{fig:fig2} Cross view of periodic metallic structure
with scattering matrices for the existing interfaces.}
\end{figure}
\begin{align}
&H_{z}^{(1)}(x,y)=\sum^{+\infty}_{n=-\infty}\left(a_{n}^{(1)}(y)e^{-jK_{y,n}^{(1)}y}+b_{n}^{(1)}(y)
\right.
\nonumber \\
&\left.e^{jK_{y,n}^{(1)}y}\right)\frac{e^{-j\frac{2\pi
n}{d_x}x}}{\sqrt{d_x}} \label{eq:one},
\end{align}
\begin{align}
H_{z}^{(2)}(x,y)=\sum^{+\infty}_{n=0}\left(a_{n}^{(2)}(y)e^{-jK_{y,n}^{(2)}y}+b_{n}^{(2)}(y)e^{jK_{y,n}^{(2)}y}\right)
\nonumber \\ \sqrt{\frac{2}{{d_{x}-w_{x}}}} \cos\left(\frac{n\pi
}{d_{x}-w_{x}}\left({x-\frac{w_x}{2}}\right)\right) \label{eq:two},
\end{align}
where $j=\sqrt{-1}$, and ${a_n}^{(i)} (y)$ and ${b_n}^{(i)} (y)$ are
\textit{n}th-order diffraction coefficients for incident and
reflected waves, respectively, in the region \textit{i},
\textit{i}=1, 2. Propagation constant, $ K_{y,n}^{(i)}$, along
Y-axis for \textit{n}th-order diffraction mode in region \textit{i}
is
\begin{equation}
K_{y,n}^{(1)}=\sqrt{K_0^{2}-{\left(\frac{2n\pi}{d_x}\right)}^{2}}
 \label{eq:three},
\end{equation}
for region \textit{i}=1, and
\begin{equation}
K_{y,n}^{(2)}=\sqrt{K_0^{2}\epsilon_s-{\left(K_{x0}+\frac{n\pi}{d_x-w_x}\right)}^{2}}
 \label{eq:four},
\end{equation}
for region \textit{i}=2. ${\epsilon}_s$ is the relative permittivity
constant of dielectric in regions 2, 3 and 4, and $K_0$ is
free-space wavenumber. $K_{x0}$ in Eq.~(\ref{eq:four}) is the
initial propagation constant along X-axis. For the special case of
normally incident TM-polarized wave, $K_{x0}$ is zero, as will be
assumed in this study. It is noted that to propagate
\textit{n}th-order diffraction mode in regions \textit{i}=1 and 2,
its propagation constant $K_{y,n}^{(i)}$ along Y-direction must be
real. Modes with pure imaginary values of $K_{y,n}^{(i)}$ decay
rapidly by distance from y=0. Based on Eqs.~(\ref{eq:three}) and
(\ref{eq:four}), at high enough wavelength values ${\lambda}$ (low
enough values of $K_0$), all values of propagation constant
$K_{y,n}^{(i)}$ along Y-axis in regions \textit{i}=1 and 2 are pure
imaginary except for fundamental mode, \textit{n}=0, that
$K_{y,0}^{(i)}$ is real. In fact, all non-zero orders ($n\neq0$) of
diffraction modes are evanescent in regions 1 and 2. Therefore from
here on we will focus on the fundamental mode only. By substituting
Eqs.~(\ref{eq:one}) and (\ref{eq:two}) into the Maxwell's equation,
${\nabla}\times\overrightarrow{H}=j\epsilon\omega\overrightarrow{E}$,
tangential electric field components $E_x^{(i)}$ in regions
\textit{i}=1 and 2 are calculated as
\begin{align}
E_{x}^{(1)}(x,y)=\frac{1}{j\omega\epsilon_0}\sum^{+\infty}_{n=-\infty}(jK_{y,n}^{(1)})\left(-a_{n}^{(1)}(y)e^{-jK_{y,n}^{(1)}y}\right.\nonumber \\
\left. +b_{n}^{(1)}(y)e^{jK_{y,n}^{(1)}y}\right)
\frac{1}{\sqrt{d_x}}e^{-j\frac{2\pi n}{d_x}x} \label{eq:five},
\end{align}
\begin{align}
E_{x}^{(2)}(x,y)=\frac{1}{j\omega\epsilon_s}\sum^{+\infty}_{n=0}(jK_{y,n}^{(2)})\left(-a_{n}^{(2)}(y)e^{-jK_{y,n}^{(2)}y}
\right.\nonumber
\\ \left. +b_{n}^{(2)}(y)
e^{jK_{y,n}^{(2)}y}\right)\sqrt{\frac{2}{{d_{x}-w_{x}}}}\cos\left(\frac{n\pi
}{d_{x}-w_{x}}\left({x-\frac{w_x}{2}}\right)\right)
 \label{eq:six}.
\end{align}
In order to obtain coefficient $b_n^{(i)} (y)$ in regions
\textit{i}=1 and 2, continuity equations for tangential electric and
magnetic components at the interface of regions 1 and 2 (y=0) are
used:
\begin{subequations}
\label{eq:seven}
\begin{equation}
H_z^{(1)}(x,0)=H_z^{(2)}(x,0),~\frac{w_x}{2}<x<d_x-\frac{w_x}{2}
 ,\label{subeq:1}
\end{equation}
\begin{align}
&E_x^{(1)}(x,0) = \begin{cases} 0,0\leq x\leq{\frac{w_x}{2}},
\text{or},~d_x-\frac{w_x}{2}\leq
x\leq d_x,\\
E_x^{(2)}(x,0),\frac{w_x}{2}<x<d_x-\frac{w_x}{2} \label{subeq:2}.
\end{cases}\end{align}
\end{subequations}
By considering $H_{z}^{(1)}$ and $H_z^{(2)}$ in the forms of
Eqs.~(\ref{eq:one}) and (\ref{eq:two}), respectively, boundary
condition (\ref{subeq:1}) can be written as
\begin{align}
\frac{1}{\sqrt{d_x}}\sum^{+\infty}_{n=-\infty}\left(a_{n}^{(1)}(0)+b_{n}^{(1)}(0)\right)e^{-j\frac{2\pi
n}{d_x}x}=\sqrt{\frac{2}{d_x-w_x}}\nonumber
\\\sum^{+\infty}_{n=0}\left(a_{n}^{(2)}(0)+b_{n}^{(2)}(0)\right)\cos\left(\frac{n\pi
}{d_{x}-w_{x}}\left({x-\frac{w_x}{2}}\right)\right)
 \label{eq:eight}.
\end{align}
Similarly, boundary condition (\ref{subeq:2}) can be written as
below by expanding the electric field, $E_x$, according to
Eqs.~(\ref{eq:five}) and (\ref{eq:six}):
\begin{align}
\frac{1}{j\omega\epsilon_0}\frac{1}{\sqrt{d_x}}\sum^{+\infty}_{n=-\infty}(jK_{y,n}^{(1)})\left(-a_{n}^{(1)}(0)
+b_{n}^{(1)}(0)\right)e^{-j\frac{2\pi n}{d_x}x}\nonumber
\\
=\frac{1}{j\omega\epsilon_s}\sqrt{\frac{2}{{d_{x}-w_{x}}}}\sum^{+\infty}_{n=0}(jK_{y,n}^{(2)})\left(-a_{n}^{(2)}(0)
+b_{n}^{(2)}(0)\right)\nonumber
\\\cos\left(\frac{n\pi
}{d_{x}-w_{x}}\left({x-\frac{w_x}{2}}\right)\right)
 \label{eq:nine}.
\end{align}
Multiplying both sides of Eqs.~(\ref{eq:eight}) and (\ref{eq:nine})
by
$\cos\left(\frac{m\pi}{d_{x}-w_{x}}\left(x-\frac{w_x}{2}\right)\right)$
and integrating over a period from $x=0$ to $x=d_x$ would lead to
the following system of linear algebraic equations for coefficients
$a_n^{(1)} (0)$, $a_n^{(2)} (0)$, $b_n^{(1)} (0)$ and $b_n^{(2)}
(0)$:
\begin{align}
&a_{m}^{(2)}(0)+b_{m}^{(2)}(0)={(d_{x}-w_{x})}^{\frac{5}{2}}\sum^{+\infty}_{n=-\infty}-\frac{j\sqrt{d_{x}}e^{-\frac{jn\pi(2d_{x}+w_{x})}
{d_{x}}}n}{\pi\left(d_{x}(m+2n)-2nw_{x}\right)}\nonumber
\\
&\times \frac{\left(-1+{(-1)}^m e^{\frac{2jn\pi
w_x}{d_x}}\right)}{\left({d_{x}}(m-2n)+2nw_{x}\right)}
 \left(a_{n}^{(1)}(0)+b_{n}^{(1)}(0)\right)
\label{eq:ten},\end{align}
\begin{align}
&\frac{K_{y,m}^{(2)}\epsilon_{0}}{K_{y,m}^{(1)}\epsilon_{s}}\left(a_{m}^{(2)}(0)-b_{m}^{(2)}(0)\right)=\sqrt{d_{x}}{(d_x-w_x)}^{\frac{3}{2}}\nonumber
\\
&\sum^{+\infty}_{n=-\infty}
\left(\frac{2jn(d_x-w_x)\left(\cos{\left(\frac{m\pi
w_x}{2(d_{x}-w_x)}\right)}-\cos{\left(\frac{m\pi
(-2d_x+w_x)}{2(d_{x}-w_x)}\right)}\right)}
{2\pi\left(d_{x}(m+2n)-2nw_{x}\right)
\left({d_{x}}(m-2n)+2nw_{x}\right)}\right.\nonumber
\\
&\left.+\frac{{d_x}m \left(\sin{\left(\frac{m\pi
(-2d_x+w_x)}{2(d_{x}-w_x)}\right)}+\sin{\left(\frac{m\pi(w_x)}{2(d_{x}-w_x)}\right)}\right)}{2\pi\left(d_{x}(m+2n)-2nw_{x}\right)
\left({d_{x}}(m-2n)+2nw_{x}\right)}\right)\nonumber
\\& \left(b_{n}^{(1)}(0)-a_{n}^{(1)}(0)\right)
 \label{eq:eleven}.
\end{align}
As mentioned in the followings of Eq.~(\ref{eq:four}), at high
enough wavelength values, $K_{y,n}^{(1)}$ and $K_{y,n}^{(2)}$ for
$n\neq0$ are pure imaginary; and so, they decay rapidly in
Y-direction distance from y=0. Therefore, incident and reflected
coefficients, $a_0^{(1)}(y)$, $a_0^{(2)}(y)$, $b_0^{(1)}(y)$,
$b_0^{(2)}(y)$, of guided order diffraction mode \textit{m}=0 and
\textit{n}=0, are considered in Eqs.~(\ref{eq:ten}) and
(\ref{eq:eleven}) only. By substituting \textit{m}=0 and
\textit{n}=0, coefficients of zeroth-order diffraction mode are
calculated in matrix form as
\begin{eqnarray}
 \left (
\begin{array}{c}
b_0^{(1)}(0) \\b_0^{(2)}(0)
\end{array}\right)
=\bm{S}^{(1-2)}\left (
\begin{array}{c} a_0^{(1)}(0) \\b_0^{(2)}(0)
\end{array}\right)
\label{eq:twelve},\end{eqnarray} with
\begin{equation}
\bm{S}^{(1-2)} =\left(
\begin{array}{cc}
 1+\frac{2(-d_x+w_x)}{d_x-w_x+d_x\sqrt{\epsilon_s}} &  \frac{2\sqrt{d_x}\sqrt{d_x-w_x}\sqrt{\epsilon_s}}{d_x-w_x+d{_x}\epsilon_s}\\ \frac{2\sqrt{d_x}\sqrt{d_x-w_x}\sqrt{\epsilon_s}}{d_x-w_x+d{_x}\epsilon_s} & -1+\frac{2(d_x-w_x)}{d_x-w_x+d_x\sqrt{\epsilon_s}}
\end{array}
\right) \label{eq:thrteen}.
\end{equation}
Reflection coefficient $b_0^{(1)}\left(-w_y/4\right)$ at $y=-w_y/4$
and transmission coefficient $b_0^{(2)}\left(w_y/4\right)$ at
$y=w_y/4$ can be obtained from:
    \begin{align}
&\left (
\begin{array}{c}
b_0^{(1)}\left(-\frac{w_y}{4}\right)
\\b_0^{(2)}\left(\frac{w_y}{4}\right)
\end{array}\right)
=\bm{T}^{(1-2)}\bm{S}^{(1-2)}\bm{T}^{(1-2)}\left (
\begin{array}{c} a_0^{(1)}\left(-\frac{w_y}{4}\right) \\a_0^{(2)}\left(\frac{w_y}{4}\right)
\end{array}\right)
\label{eq:fourteen},
\end{align}
where $\bm{T}^{(1-2)}$ is the transferring matrix:
    \begin{equation}
\bm{T}^{(1-2)}= \left(
\begin{array}{cc}
 e^{-jK_{0}\frac{w_y}{4}} &  0\\ 0 & e^{-jK_{0}\sqrt{\epsilon_s}\frac{w_y}{4}}
\end{array}
\right) \label{eq:fifteen}.
\end{equation}

\subsection{Calculation of $\bm{S}^{(2-3)}$ Matrix}
\label{sec:3} In order to calculate the scattering matrix at the
interface of regions 2 and 3, $\bm{S}^{(2-3)}$, magnetic fields in
the upper and lower sides of PPWGs (in the middle of stepped-slit)
interface are required. Similar to Fourier series expansion Eq.
(\ref{eq:two}) which was valid for region 2, magnetic field in
region 3 can be expanded as
\begin{align}
H_{z}^{(3)}(x,y)=\sum^{+\infty}_{n=0}\left(-a_{n}^{(3)}(y)e^{-jK_{y,n}^{(3)}y}-b_{n}^{(3)}(y)
e^{jK_{y,n}^{(3)}y}\right)\nonumber
\\ \sqrt{\frac{2}{{d_{x}-w^{'}_{x}}}}
\cos\left(\frac{n\pi
}{d_{x}-w_{x}^{'}}\left({x-\frac{w^{'}_x}{2}}\right)\right)
 \label{eq:sixteen},\end{align}
where the minus sign comes to be consistent with the convention of
Fig. \ref{fig:fig2}. $K_{y,n}^{(3)}$ is the propagation constant
along Y-direction for \textit{n}th-order diffraction mode in region
3 written as
\begin{equation}
K_{y,n}^{(3)}=\sqrt{K_0^{2}\epsilon_s-{\left(\frac{n\pi}{d_x-w^{'}_{x}}\right)}^{2}}
 \label{eq:seventeen}.
\end{equation}
To obtain the x-component of electric field in region 3,
$E_x^{(3)}$, a similar procedure as the one that led to
Eq.~(\ref{eq:six}) should be carried out, which will give
\begin{align}
E_{x}^{(3)}(x,y)=\frac{1}{j\omega\epsilon_s}\sum^{+\infty}_{n=0}(jK_{y,n}^{(3)})\left(a_{n}^{(3)}(y)e^{-jK_{y,n}^{(3)}y}
-b_{n}^{(3)}(y)\right. \nonumber
\\ \left.e^{jK_{y,n}^{(3)}y}\right)\sqrt{\frac{2}{{d_{x}-w^{'}_{x}}}}\cos\left(\frac{n\pi
}{d_{x}-w^{'}_{x}}\left({x-\frac{w^{'}_x}{2}}\right)\right)
 \label{eq:eighteen}.
\end{align}
Reflection coefficients $b_n^{(i)}(y)$, \textit{i}=2, 3, can be
calculated from the boundary conditions at the interface of regions
2 and 3; i.e. $H_z$ and $E_x$ should vary continuously across $y=
w_y/2$:
\begin{subequations}
\label{eq:nineteen}
\begin{equation}
H_z^{(2)}(x,w_{y}/2)=H_z^{(3)}(x,w_{y}/2),~w_x^{'}/2<x<d_x-w_x^{'}/2
 ,\label{subeq:3}
\end{equation}
\begin{align}
E_x^{(2)}\left(x,w_{y}/2\right)=
\begin{cases}
0,0\leq x\leq w_x^{'}/2, \text{or}, d_x-w_x^{'}/2\leq x\leq d_x, \\
E_x^{(3)}\left(x,{w_y}/2\right),w_x^{'}/2<x<d_x-w_x^{'}/2.
\end{cases}
\label{subeq:4}
\end{align}\end{subequations}
By substituting Eqs.~(\ref{eq:two}), (\ref{eq:six}),
(\ref{eq:sixteen}) and (\ref{eq:eighteen}) into the
Eqs.~(\ref{subeq:3}) and (\ref{subeq:4}), and multiplying both sides
by
$\cos\left(\frac{m\pi}{d_{x}-w^{'}_{x}}\left(x-\frac{w^{'}_{x}}{2}\right)\right)$
and subsequently integrating over the slot of bottom PPWG
$(\frac{w_{x}^{'}}{2}<x<d_x-\frac{w_{x}^{'}}{2})$, the following
system of algebraic equations will be obtained for reflection and
transmission coefficients $b_0^{(2)} (w_y/2)$ and $b_0^{(3)}
(w_y/2)$ for the fundamental mode:
\begin{eqnarray}
 \left (
\begin{array}{c}
b_0^{(2)}(w_{y}/2) \\b_0^{(3)}(w_{y}/2)
\end{array}\right)
=\bm{S}^{(2-3)}\left (
\begin{array}{c} a_0^{(2)}(w_{y}/2) \\a_0^{(3)}(w_{y}/2)
\end{array}\right)
\label{eq:twenty},
\end{eqnarray}
where $\bm{S}^{(2-3)}$ is
 \begin{equation}
\bm{S}^{(2-3)}= \left(
\begin{array}{cc}
\frac{(w_{x}-w^{'}_x)}{-2d_x+w_x+w^{'}_x} &  -\frac{2\sqrt{d_x-w_{x}}\sqrt{d_x-w^{'}_x}}{-2d_x+w_x+w^{'}_{x}}\\
-\frac{2\sqrt{d_x-w_{x}}\sqrt{d_x-w^{'}_{x}}}{-2d_{x}+w_{x}+w^{'}_{x}}
& \frac{(-w_{x}+w^{'}_{x})}{-2d_x+w_x+w^{'}_x}
\end{array}
\right) \label{eq:twentyone}.
\end{equation}
Reflection coefficient $b_0^{(2)}\left(w_y/4\right)$ and
transmission coefficient $b_0^{(3)}\left(3w_y/4\right)$ at
respectively $ y=w_y/4$ and $y=3w_y/4$ can be obtained from
\begin{align}
&\left (
\begin{array}{c}
b_0^{(2)}\left(\frac{w_{y}}{4}\right)
\\b_0^{(3)}\left(\frac{3w_{y}}{4}\right)
\end{array}\right)
=\bm{T}^{(2-3)}\bm{S}^{(2-3)}\bm{T}^{(2-3)}\left (
\begin{array}{c} a_0^{(2)}\left(\frac{w_y}{4}\right) \\a_0^{(3)}\left(\frac{3w_{y}}{4}\right)
\end{array}\right)
\label{eq:twentytwo},
\end{align}
where the transferring matrix
\begin{equation}
\bm{T}^{(2-3)}= \left(
\begin{array}{cc}
 e^{-jK_{0}\sqrt{\epsilon_s}\frac{w_{y}}{4}} &  0\\ 0 &
 e^{-jK_{0}\sqrt{\epsilon_s}\frac{3w_{y}}{4}}
\end{array}
\right) \label{eq:twentythree},
\end{equation}
describes the relation between $b_0^{(2)} (y)$ at $y=w_{y}/2$ and
$y=w_{y}/4$ as well as $b_0^{(3)}(y)$ at $y=w_{y}/2$ and
$y=3w_{y}/4$.

\subsection{Calculation of $\bm{S}^{(3-4)}$ Matrix}
\label{sec:4}

The scattering matrix $\bm{S}^{(3-4)}$ at the interface of regions 3
and 4, can be calculated in a similar fashion as what we did in
Section \ref{sec:2} for calculating $\bm{S}^{(1-2)}$. Note that here
region 4 is the substrate with relative permittivity constant,
$\epsilon_s$.

\subsection{Calculation of Total Scattering Matrix, $\bm{S}^{(1-4)}$ }
\label{sec:5} The total scattering matrix $\bm{S}^{(1-4)}$ can be
calculated by successive cascade of transferred interface scattering
from
\begin{align}
&\bm{T}^{(1-4)}\bm{S}^{(1-4)}\bm{T}^{(1-4)}=\text{Cas}\left\{\text{Cas}\left\{\bm{T}^{(1-2)}\bm{S}^{(1-2)}\bm{T}^{(1-2)},\right.\right.
\nonumber
\\
&\left.\bm{T}^{(2-3)}\bm{S}^{(2-3)}\bm{T}^{(2-3)}\right\},
\left.\bm{T}^{(3-4)}\bm{S}^{(3-4)}\bm{T}^{(3-4)}\right\}
\label{eq:twentyfour},
\end{align}
where here-defined operator Cas\{A, B\} operates on two by two
matrices A and B as (\cite{Pozar})
\begin{equation}
\text{Cas}\{A,B\}=\left(
\begin{array}{cc}
A_{11}+\frac{A_{12}B_{11}A_{21}}{1-B_{11}A_{22}} &
\frac{A_{12}B_{12}}{1-B_{11}A_{22}}\\
\frac{B_{21}A_{21}}{1-A_{22}B_{11}} &
B_{22}+\frac{B_{21}A_{22}B_{12}}{1-A_{22}B_{11}}
\end{array}
\right) \label{eq:twentyfive}.
\end{equation}
Subsequently, the total reflection coefficient $b_0^{(1)}(0)$ at y=0
and total transmission coefficient $b_0^{(4)} (w_y)$ at $y=w_y$ can
be obtained as
\begin{eqnarray}
\left (
\begin{array}{c}
b_0^{(1)}(0) \\b_0^{(4)}(w_{y})
\end{array}\right)
= \bm{S}^{(1-4)}\left (
\begin{array}{c} a_0^{(1)}(0) \\a_0^{(4)}(w_{y})
\end{array}\right)
\label{twentysix}.
\end{eqnarray}
Here, for the sake of brevity, only the expression for $b_0^{(1)}
(0)$ is presented
\begin{subequations}
\label{eq:twentyseven}
\begin{equation}
b_0^{(1)}(0)=e^{jK_0
w_y}\left(\frac{1}{-1+c-\sqrt{\epsilon_s}}+\frac{\Psi_1}{\Psi_2}\right)a_0^{(1)}(0),
\label{subeq:5}
\end{equation}
\text{with}
\begin{align}
&\Psi_1=-(-2+c^{'})(-2+c+c^{'})e^{4jK_{0}w_{y}\sqrt{\epsilon_s}}\left((-1+c)c-\sqrt{\epsilon_s}\right.
\nonumber \\
&\left.-\epsilon_s\right)+
c^{'}(-2+c+c^{'})\left((-1+c)c+(3-2c)\sqrt{\epsilon_s}+\epsilon_s\right)\nonumber \\
&-2(c-c^{'})e^{2jK_{0}w_{y}\sqrt{\epsilon_s}}\left((-1+c)c+
\left(3+c(-2+c^{'})-2c^{'}\right)\right. \nonumber
\\ &\left. \sqrt{\epsilon_s}+\epsilon_s-c^{'}\epsilon_s\right),\label{subeq:6}
\end{align}
\begin{align}
&\Psi_2=(-2+c^{'})(-2+c+c^{'})e^{4jK_{0}w_{y}\sqrt{\epsilon_s}}(1-c+\sqrt{\epsilon_s})^{2}+\nonumber
\\ &2(c+c^{'})(-2+c+c^{'})
e^{2jK_{0}w_{y}\sqrt{\epsilon_s}}
\left(-1+c-(-1+c^{'})\sqrt{\epsilon_s}\right)\nonumber
\\ & -c^{'}(-2+c+c^{'})\left({(-1+c)}^{2}-\epsilon_s\right)\label{subeq:7},
\end{align}
\end{subequations}
where $c =w_x/d_x$ and $c' =w'_x/d_x$ are, respectively, the
normalized widths of top and bottom metallic rods.

\section{Results And Discussion}
\label{sec:6}
\subsection{Design Method}
\label{sec:7} 
\begin{figure}
\includegraphics[scale=.5]{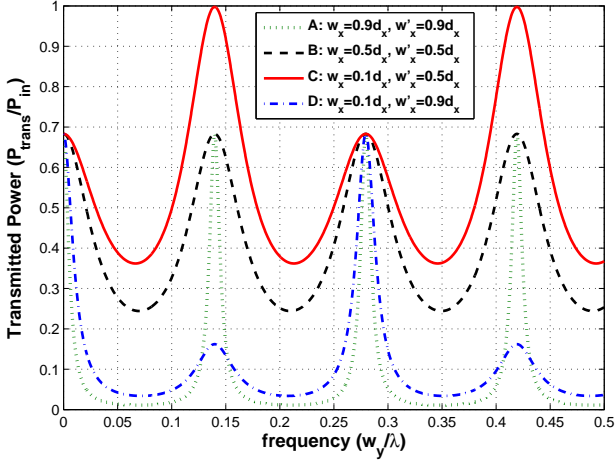}
\caption{\label{fig:fig3} Power of TM-incident wave transmitted
through the structure, as a function of frequency for structures A
to D with different slit widths. The curves have been calculated by
the theoretical model proposed in this work, assuming relative
dielectric permittivity as 11.7 and absorption coefficient as 8000
$\text{cm}^{-1}$.}
\end{figure}

Analytical expression for the reflection coefficient
$b_0^{(1)}(0)$ of zeroth-order diffraction mode for the impinging
TM-polarized wave on the structure was calculated in the last
section as Eq.~(\ref{subeq:5}). The power $1-|b_0^{(1)}(0)|^2$
transmitted into the substrate, depends on the wavelength of the
incident wave, periodicity $d_x$ of the structure, width $w_x$ of
the top metallic rod, width $ w'_x$ of the bottom metallic rod,
height $w_y$ of stepped-slit, and relative permittivity constant
$\epsilon_s$ of the dielectric of structure. In this section, the
dielectric is assumed to be $
\text{In}_{0.53}\text{Ga}_{0.47}\text{As}$, with relative
permittivity constant 11.7 (\cite{Takagi}) and absorption
coefficient 8000 $\text{cm}^{-1}$ \cite{Carpintero}. The ultimate
goal of the design is to achieve a structure with maximum
transmitted power of incident TM-polarized wave at a specific
optical frequency and all the range of terahertz frequency
bandwidth, by determining the associated parameters.
Fig.~\ref{fig:fig3} shows the transmitted power calculated by our
theoretical model with respect to height of the proposed structure
$w_y/\lambda$ (normalized to wavelength) at several widths $w_x$ and
$w_{x}^{'}$ of metallic rods; different curves correspond to
different values of $w_x$ and $w_{x}^{'}$. Several resonant guided
modes can be observed in the figure. However, first guided mode, at
$w_y/\lambda$ close to zero, is a non-resonant mode. At this mode,
the wavelength of the impinging wave is much more than the height
$w_y$ of structure, and metallic rods widths have no effect on the
power transmitted into the substrate. Maximum transmission power at
this mode is
\begin{equation}
\bm{T}_{\text{trans}}=\frac{4\sqrt{\epsilon_s}}{{(1+\sqrt{\epsilon_s})}^{2}}
\label{eq:twentyeight}.
\end{equation}
Frequency values of the resonant modes in Fig.~\ref{fig:fig3} agree
well with $\lambda_n=\frac{2 w_y}{n}\sqrt{\epsilon_{s}}~( n=1,2,
\ldots)$ proposed by the reference \cite{Hsieh}. However for certain
ratios of $w_x$ and $w^{'}_{x}$ (for example $w_{x}^{'}/{w_x}=5$
shown by red curve) transmitted power with much larger values than
those reported in \cite{Hsieh} for uniform slit structure, can be
obtained at certain frequencies. This is an interesting property of
stepped periodic metallic structures, and will be discussed further
later. It is also noted that frequency values of resonant guided
modes depend on $w_y$ only, whereas transmitted power depends on
solely $w_x$ and $w^{'}_{x}$. Therefore at the first step of design,
the height $w_y$ of the structure is determined, based on the
optical frequency desired to transmit maximum electromagnetic power.
Maximum power transmission is read at $w_y/\lambda=0.14$ from
Fig.~\ref{fig:fig3}.

Field intensity, \textit {I} in unit depth of the substrate can be
calculated from
\begin{equation}
{I}=\frac{{P}_{\text{trans}}}{d_x-w^{'}_{x}}, \label{eq:twentynine}
\end{equation}
where $P_{\text{trans}}$ is the power transmitted into the
substrate, and $d_x-w^{'}_{x}$ is the bottom slit width. According
to the above equation \textit{I} increases with the increase of
$w^{'}_{x}$. Fig.~\ref{fig:fig3} contains $P_{\text{trans}}$ for
uniform-slit structure, A and B, as well as stepped-slit structures
C and D. First comparing two uniform-slit structures A and B, since
$w^{'(\text{A})}_{x}>w^{'(\text{B})}_{x}$ and
$P_{\text{trans}}^{(\text{A})}=P_{\text{trans}}^{(\text{B})}$ at
$w_y/\lambda=0.28$, so $I^{(\text{A})}>I^{(\text{B})}$. Frequency
bandwidth of A though has a smaller value than that of B. Therefore,
in metallic arrays of uniform slits, field intensity \textit{I} and
frequency bandwidth have opposite trends of change, such that
simultaneous enhancement of both quantities field intensity and
bandwidth is impossible, as has been also pointed out by
\cite{Hsieh}. Second comparing stepped-slit structure D with
uniform-slit structure A,
$P_{\text{trans}}^{(\text{A})}=P_{\text{trans}}^{(\text{D})}=0.7 $
at $w_{y}/\lambda=0.28$, and
$w^{'(\text{A})}_{x}=w^{'(\text{D})}_{x}$, therefore
$I^{(\text{A})}=I^{(\text{D})}$. At the same time, frequency
bandwidth of D is larger than that of A. As a result, as opposed to
uniform-slits arrays, interestingly metallic arrays with
stepped-slits can be utilized for increasing the frequency
bandwidth, while preserving field intensity \textit{I} in the
substrate constant.

\begin{figure}
\includegraphics[scale=.5]{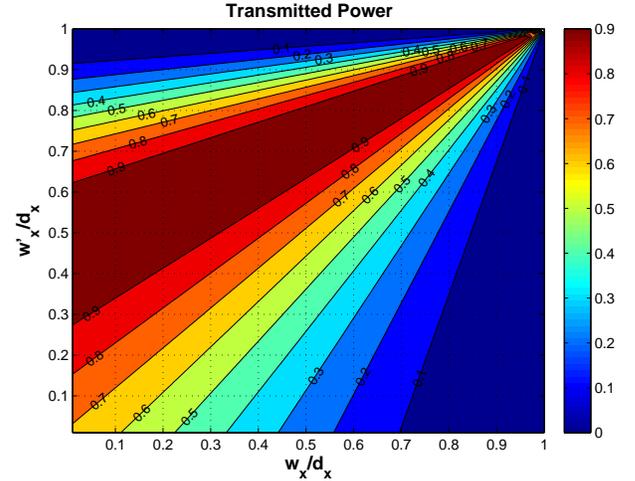}
\caption{\label{fig:fig4} Transmitted power of TM-polarized incident
wave versus normalized widths of top and bottom metallic rods
$(w_{x}/d_{x}$ and $w^{'}_{x}/d_{x}$) for $w_y/\lambda=0.14$,
$\epsilon_s=11.7$ and absorption coefficient 8000 $\text{cm}^{-1}$
(Maximum transmitted power is $99\%$).}
\end{figure}

\begin{figure}
\includegraphics[scale=.55]{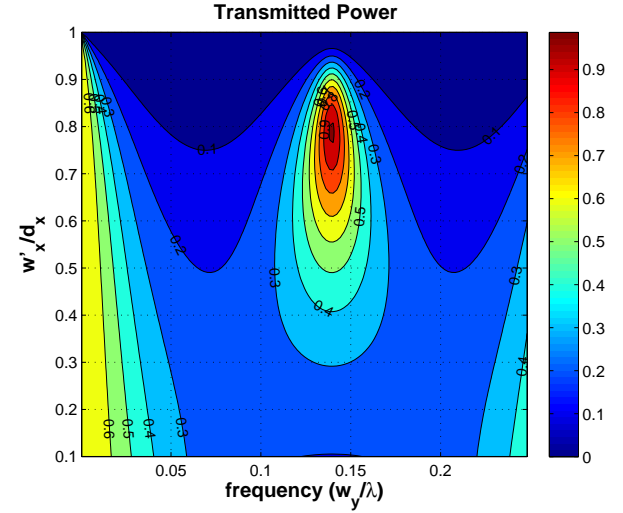}
\caption{\label{fig:fig5} Transmitted power of TM-polarized incident
wave versus frequency $(w_{y}/\lambda)$, (structure height
normalized to wavelength) and normalized width of bottom metallic
rod to the periodicity of the proposed structure ($w^{'}_{x}/d_x$)
for $w_{x}/d_x=0.6$ (Maximum transmitted power is $99\%$).}
\end{figure}

In the second step of design, suitable widths $w_x$ and $w^{'}_{x}$
should be determined. To this end, transmitted power of TM-incident
wave has been calculated from Eq.~(\ref{subeq:5}), and is plotted in
Fig.~\ref{fig:fig4} as a function of $w_x/d_x$ and $w^{'}_{x}/d_x$,
for $K_0 w_y=2\pi (0.14)$, $\epsilon_s=11.7$ and absorption
coefficient 8000 $\text{cm}^{-1}$. The equi-level surfaces are
triangles, and the one corresponding to maximum power transmission
of 99\% is shown by dark red color with the base at $w_x/d_x =0$.
Note that the special case of $w_x=w^{'}_{x}$ reduces the stepped
slit to uniform slit with maximum power transmission of 70\%.
Therefore, interestingly, almost entire power transmission is
possible for metallic array of stepped slits, which was not the case
for uniform slits.
Here we assume $ w_x/d_x =0.6<1$. Fig.~\ref{fig:fig5} demonstrates
transmitted power of TM-incident wave as a function of $w_y/\lambda$
and $w^{'}_{x}/d_x$. Power transmission as high as even 99\% is
possible for certain values of $w^{'}_{x}/d_x$. Moreover, the
magnitude of power transmissivity at THz frequency (non-resonant
mode) in Fig.~\ref{fig:fig5} is independent of metallic rods widths
($w_x$ and $w^{'}_{x}$) and its value is about 70\%; this fact was
also observed in Fig.~\ref{fig:fig3} for small values of
$w_y/\lambda$. According to Fig.~\ref{fig:fig5}, at
$w_y/\lambda=0.14$, maximum power transmission of 99\% corresponds
to $w^{'}_{x}/d_x=0.8$.

\subsection{Results of the Designed Structure}
\label{sec:8} 
\begin{figure}
\includegraphics[scale=.5]{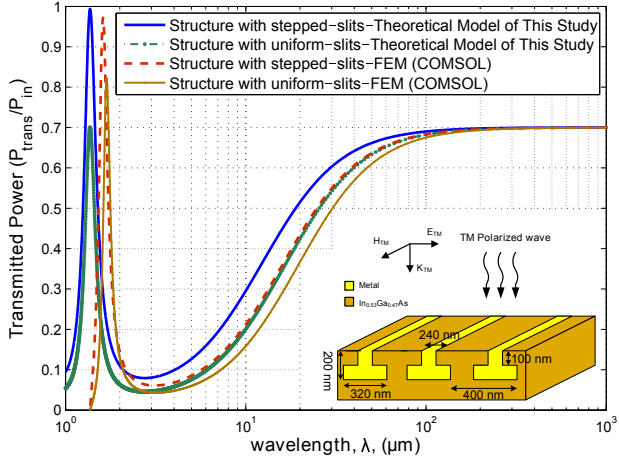}
\caption{\label{fig:fig6} Transmitted power of TM-incident wave
through proposed structure (stepped-slit) and uniform-slit periodic
structure versus wavelength, obtained by the presented theoretical
and FEM models.}
\end{figure}

In the previous section, normalized structure
parameters were designed to be $w_y/\lambda=0.14$, $w_x/d_x =0.6$,
and $w^{'}_{x}/d_x =0.8$. For the sake of comparison, the same
values for $\lambda$ and $d_x$ are adopted here in the analytical
and numerical models as those of reference \cite{Hsieh};
$\lambda$=1.5 \text{$\mu$m} and $d_{x}$=400 nm, leading to
$w_{y}$=200 nm, $w_{x}$=240 nm, and $w^{'}_{x}$=320 nm.
Fig.~\ref{fig:fig6} shows the designed structure as well as its
transmitted power calculated both analytically using
Eq.~(\ref{subeq:5}) and numerically by the finite element method
(FEM) within COMSOL package.
\begin{figure}
\centering
\subfigure[]{\label{fig:f71}\includegraphics[scale=.5]{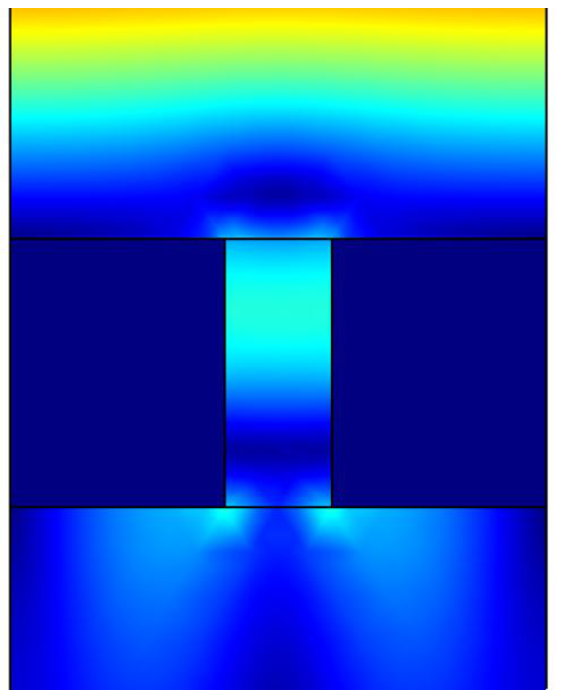}}
\centering
\subfigure[]{\label{fig:f72}\includegraphics[scale=.5]{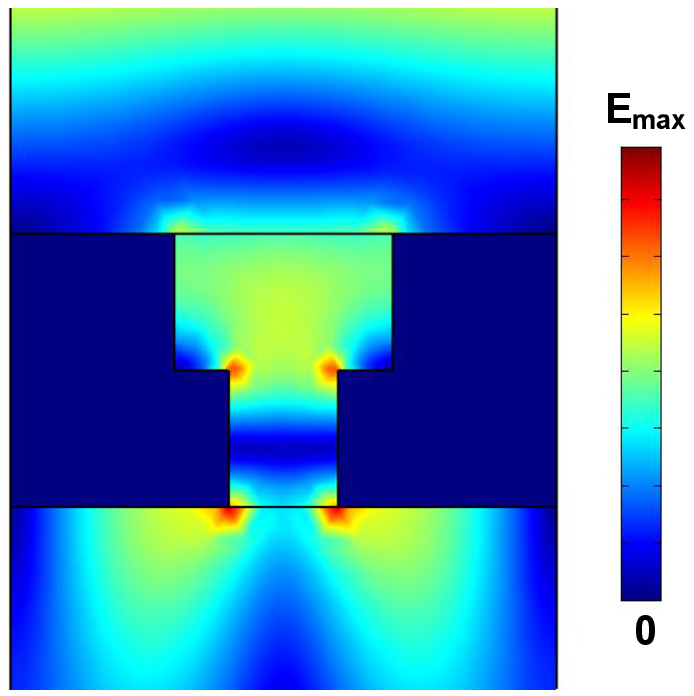}}
\centering \caption{Electric field distribution in a) uniform-slit
structure with $w_y$=200 nm, $w_x$=320 nm, and b) the designed
structure (stepped-slits) with $w_y$=200 nm, $w_x$=240 nm,
$w^{'}_x$=320 nm, at $\lambda$=1.5 \text{$\mu$m}.} \label{fig:fig7}
\end{figure}
Transmitted power for a uniform-slits array having similar
dimensions $d_x$=400 nm, $w_y$=200 nm as the stepped-slits array but
with $w_x$=320 nm has been also plotted in Fig.~\ref{fig:fig6} for
comparison purpose. Analytical results of Fig.~\ref{fig:fig6} show
good agreements with FEM results for both stepped-slits and
uniform-slits arrays. The slight difference between resonance
wavelength $\lambda$ of green and brown curves corresponding to
uniform-slits array, can be justified as follows, in contrast to
which considered in the proposed theoretical model, the guided mode
of PPWGs slightly penetrates into air ($y=0^-$) and substrate
($y=w_{y}^+$). This fact can be think of as considering the
effective value of $w_y$ in the FEM model to be larger than the
value of $w_y$ in the analytical model, i.e.
$w_{y}^{\text{(fem)}}>w_{y}^{\text{(analytic)}}(=w_y)$. On the other
hand, according to Fig.~\ref{fig:fig3}, at first resonance mode
$w_{y}^{\text{(analytic)}}/\lambda=w_{y}^{\text{(fem)}}/\lambda=0.14$.
Therefore, $\lambda^{\text{(fem)}}>\lambda^{\text{(analytic)}}$,
which justifies the difference between analytic and FEM resonance
wavelengths of Fig.~\ref{fig:fig6} for uniform-slits arrays. A
similar reasoning applies to the difference between resonance
wavelength of blue and red curves associated with stepped-slits
arrays. As for comparison between performance of stepped- and
uniform-slits arrays, maximum transmitted power of stepped-slits
array reaches to the impressive value of about 100\% at
$\lambda$=1.5-1.6 \text{$\mu$m}. For the case of uniform-slits array
this value is only 70\%. This fact is also confirmed by Fig.
~\ref{fig:fig7} which shows that the value of electric field inside
the substrate of stepped-slits arrays is larger than that inside
uniform-slits arrays. At terahertz frequency band though, both
stepped- and uniform-slits arrays have maximum transmitted power of
70\%. Therefore, the designed stepped-slits array can transmit
entire power of incident wave at optical frequency, while
simultaneously transmitting 70\% at terahertz frequency band.

\section{Conclusion}
\label{sec:9} Periodic metallic structures have been of both
research interest and practical implication for power transmission
at optical frequencies. Surface waves excited by incident transverse
magnetic wave are responsible for power transmission through
structure into the substrate. In this paper, metallic structure with
periodic array of stepped-slits was suggested. A closed-form
expression for the power transmitted through the structure was
obtained. The expression takes on structure height, top and bottom
slit widths, permittivity constant of dielectric substrate, and
frequency as unknown parameters. Conversely, for given optical
wavelength corresponding to the entire power transmission,
dimensions of the structure can be designed. As an example, for
typical optical wavelength of 1.5 \text{$\mu$m}, results show that
70\% of incident power is transmitted to
$\text{In}_{0.53}\text{Ga}_{0.47}\text{As}$ substrate in terahertz
frequency band, simultaneous to the entire power transmission at the
optical frequency of design. As opposed to uniform-slits, with
stepped-slits field intensity in the substrate can be increased
while preserving frequency range of maximum power transmission
constant. All the results were validated numerically by finite
element method. For the purpose of increasing efficiency as well as
radiated terahertz power, the proposed structure can find
applications in the periodic structure of plasmonic photoconductive
antennas.


%





\ifCLASSOPTIONcaptionsoff
  \newpage
\fi

\end{document}